# Differential Fast Fixed-Point Algorithms for Underdetermined Instantaneous and Convolutive Partial Blind Source Separation

Johan Thomas, Yannick Deville, *Member, IEEE*, and Shahram Hosseini

*Abstract*—This paper concerns underdetermined linear instantaneous and convolutive blind source separation (BSS), i.e., the case when the number $P$ of observed mixed signals is lower than the number $N$ of sources. We propose partial BSS methods, which separate $P$ supposedly nonstationary sources of interest (while keeping residual components for the other $N - P$, supposedly stationary, "noise" sources). These methods are based on the general differential BSS concept that we introduced before. In the instantaneous case, the approach proposed in this paper consists of a differential extension of the FastICA method (which does not apply to underdetermined mixtures). In the convolutive case, we extend our recent time-domain fast fixed-point C-FICA algorithm to underdetermined mixtures. Both proposed approaches thus keep the attractive features of the FastICA and C-FICA methods. Our approaches are based on differential sphering processes, followed by the optimization of the differential nonnormalized kurtosis that we introduce in this paper. Experimental tests show that these differential algorithms are much more robust to noise sources than the standard FastICA and C-FICA algorithms.

## I. INTRODUCTION

**B**LIND source separation (BSS) methods [1] aim at restoring a set of $N$ unknown source signals $s_j(n)$ from a set of $P$ observed signals $x_i(n)$. The latter signals are in many cases linear instantaneous or convolutive mixtures of the source signals. Convolutive mixtures read

$$\mathbf{x}(n) = \mathbf{A}(n) * \mathbf{s}(n) \tag{1}$$

where $\mathbf{s}(n) = [s_1(n), \cdots, s_N(n)]^T$ and $\mathbf{x}(n) = [x_1(n), \cdots, x_P(n)]^T$ are the source and observation vectors, $*$ denotes the convolution operator, and the mixing matrix $\mathbf{A}(n)$ is composed of the impulse responses of unknown mixing filters. This general framework includes linear instantaneous mixtures. Then, $\mathbf{A}(n)$ reduces to a constant scalar mixing matrix $\mathbf{A}$ and the observations read

$$\mathbf{x}(n) = \mathbf{A}\mathbf{s}(n). \tag{2}$$

Here, we assume that the signals and mixing matrix are real-valued and that the sources are zero-mean and mutually statistically independent, and we propose BSS methods based on independent component analysis (ICA). Moreover, we consider the underdetermined case, i.e., $P < N$, and we require that $P \geq 2$. Various analyses and BSS methods have been reported for these difficult mixtures, most often for linear instantaneous ones [2]–[15] but also in the convolutive case [16]–[20] or even for nonlinear mixtures [21]. Many of them assume source sparsity [5], [6], [9]–[13], [15]–[18], [20], [21]. Some other methods set other restrictions on the sources [8], [13], [14], [16], [19] or on the mixing conditions [2], [19]. Discrete sources are especially considered in some cases [3], [4], [7]. Here, we aim at avoiding all these constraints.

In [22], we introduced a general differential BSS concept for processing underdetermined mixtures. In its standard version, we consider the situation when (at most) $P$ of the $N$ mixed sources are nonstationary while the other $N - P$ sources (at least) are stationary. The $P$ nonstationary sources are the signals of interest in this approach, while the $N - P$ stationary sources are considered as "noise sources." Then, our differential BSS concept achieves the "partial BSS" of the $P$ sources of interest, i.e., it yields output signals which each contain contributions from only one of these $P$ sources, still superimposed with some residual components from the noise sources (this is described in [22]). However, we point out that this partial separation can be considered as a pseudocomplete separation task in situations when one does not aim at separating the noise sources since they do not include information. This method can be of practical use for the noisy "cocktail party" scenario, when some stationary noise sources are present in addition to the speech signals to be separated. One may also use our approach in multiple-input–multiple-output (MIMO) communication systems, in which received signals are often disturbed by stationary noise sources in real applications. This method may also be applied to biomedical signals, which often contain various stationary noise components.

Although we first defined this differential BSS concept in a quite general framework in [22], then we only applied it to a simple but restrictive BSS method, which is especially limited to $P = 2$ mixtures, only involving two strictly causal moving average (MA) mixing filters (i.e., no instantaneous mixing), and based on slow-convergence algorithms. Here, we introduce much more powerful BSS criteria and associated algorithms, based on differential BSS, for both linear instantaneous and convolutive mixtures. Our instantaneous method is obtained by extending to underdetermined mixtures the kurtotic separation criterion [23] and the associated, fast converging, fixed-point, FastICA algorithm [24]. In the convolutive case, we extend to underdetermined mixtures the time-domain fast fixed-point algorithm restricted to overdetermined convolutive ICA (i.e., $P \geq N$) that we recently presented in [25]. We thus keep the attractive features of the latter algorithm.

This paper is organized as follows. In Section II, we describe our fast fixed-point algorithm for underdetermined linear instantaneous mixtures derived from the differential source separation concept. In Section III, first we summarize the principles of our previous method, that achieves fast fixed-point separation for (over)determined convolutive mixtures, and then, derive from this a differential extension that aims at achieving the partial separation of the sources of interest. Section IV consists of experimental results which compare our differential instantaneous and convolutive algorithms with the associated standard ones. Conclusions are derived from these investigations in Section V.

## II. PROPOSED DIFFERENTIAL BSS METHOD FOR LINEAR INSTANTANEOUS MIXTURES

### A. New BSS Criterion Based on Differential Kurtosis

The standard FastICA method [24], which is only applicable to linear instantaneous mixtures with $P = N$ (or $P > N$), extracts a source by means of a two-stage procedure. The first stage consists in transferring the observation vector $\mathbf{x}(n)$ through a real $P \times P$ matrix $\mathbf{B}$, which yields the vector

$$\mathbf{z}(n) = \mathbf{B}\mathbf{x}(n). \quad (3)$$

In the standard FastICA method, $\mathbf{B}$ is selected so as to sphere the observations, i.e., so as to spatially whiten and normalize them. Then, the second stage of that standard method consists in deriving an output signal $y(n)$ as a linear instantaneous combination of the signals contained by $\mathbf{z}(n)$, i.e.,

$$y(n) = \mathbf{w}^T \mathbf{z}(n) \quad (4)$$

where $\mathbf{w}$ is a vector, which is constrained so that $\|\mathbf{w}\| = 1$. This vector $\mathbf{w}$ is selected so as to optimize the (nonnormalized) kurtosis of $y(n)$, defined as its zero-lag fourth-order cumulant

$$\text{kurt}_y(n) = \text{cum}(y(n), y(n), y(n), y(n)). \quad (5)$$

Now, consider the underdetermined case, i.e., $P < N$. We again derive an output signal $y(n)$ according to (3) and (4). We aim at defining new criteria for selecting $\mathbf{B}$ and $\mathbf{w}$, in order to achieve the previously defined partial BSS of the $P$ sources of interest. To this end, we apply the general differential BSS concept that we described in [22] to the specific kurtotic criterion used in the standard FastICA method. Therefore, we consider two times $n_1$ and $n_2$, then introduce the differential (nonnormalized) kurtosis that we associate with (5) for these times. We define this parameter as

$$\text{Dkurt}_y(n_1, n_2) = \text{kurt}_y(n_2) - \text{kurt}_y(n_1). \quad (6)$$

Let us show that, whereas the standard parameter $\text{kurt}_y(n)$ depends on all sources, its differential version $\text{Dkurt}_y(n_1, n_2)$ only depends on the nonstationary sources. Equations (2)–(4) yield

$$y(n) = \mathbf{v}^T \mathbf{s}(n) \quad (7)$$

where the vector

$$\mathbf{v} = (\mathbf{B}\mathbf{A})^T \mathbf{w} \quad (8)$$

includes the effects of the mixing and separating stages. Denoting $v_q$ with $q = 1, \ldots, N$, the entries of $\mathbf{v}$, (7) implies that the output signal $y(n)$ may be expressed with respect to all sources as

$$y(n) = \sum_{q=1}^{N} v_q s_q(n). \quad (9)$$

Using cumulant properties and the assumed independence of all sources, one derives easily

$$\text{kurt}_y(n) = \sum_{q=1}^{N} v_q^4 \text{kurt}_{s_q}(n) \quad (10)$$

where $\text{kurt}_{s_q}(n)$ is the kurtosis of source $s_q(n)$, again defined according to (5). The standard output kurtosis (10), therefore, actually depends on the kurtoses of all sources. The corresponding differential output kurtosis, defined in (6), may then be expressed as

$$\text{Dkurt}_y(n_1, n_2) = \sum_{q=1}^{N} v_q^4 \text{Dkurt}_{s_q}(n_1, n_2) \quad (11)$$

where we define the differential kurtosis $\text{Dkurt}_{s_q}(n_1, n_2)$ of source $s_q(n)$ in the same way as in (6). Let us now take into account the assumption that $P$ sources are nonstationary,[1] while the other sources are stationary. By $I$, we denote the set containing the $P$ indices of the unknown nonstationary sources. The standard kurtosis $\text{kurt}_{s_q}(n)$ of any source $s_q(n)$ with $q \notin I$ then takes the same values for the times $n_1$ and $n_2$, so that $\text{Dkurt}_{s_q}(n_1, n_2) = 0$.[2] Then, (11) reduces to

$$\text{Dkurt}_y(n_1, n_2) = \sum_{q \in I} v_q^4 \text{Dkurt}_{s_q}(n_1, n_2). \quad (12)$$

This shows explicitly that this differential parameter only depends on the nonstationary sources. Moreover, for given sources and times $n_1$ and $n_2$, it may be seen as a function $f(\cdot)$ of the set of variables $\{v_q, q \in I\}$, i.e., $\text{Dkurt}_y(n_1, n_2)$ is equal to

$$f(v_q, q \in I) = \sum_{q \in I} v_q^4 \alpha_q \quad (13)$$

where the parameters $\alpha_q$ are here equal to the differential kurtoses $\text{Dkurt}_{s_q}(n_1, n_2)$ of the nonstationary sources. The type of function defined in (13) has been widely studied in the framework of standard kurtotic BSS methods, i.e., methods for the case when $P = N$, because the standard kurtosis used as a BSS criterion in that case may also be expressed according to (13).[3] The following result has been established (see [1, p. 173] for the basic two-source configuration and [23] for a general proof). Assume that all parameters $\alpha_q$ with $q \in I$ are nonzero, i.e., all nonstationary sources have nonzero differential kurtoses for the considered times $n_1$ and $n_2$. Consider the values of the function in (13) on the $P$-dimensional unit sphere, i.e., for $\{v_q, q \in I\}$ such that

$$\sum_{q \in I} v_q^2 = 1. \qquad (14)$$

---

[1]The number $\check{N}$ of nonstationary sources is assumed to be equal to the number $P$ of observations hereafter, except in Section III-E.

[2]Note that the "complete" stationarity of the sources $s_q(n)$ with $q \notin I$ is sufficient for, but not required by, our method: We only need their differential kurtoses (and their differential powers below) to be zero for the considered times. Apart from the stationarity case, the differential kurtoses are zero for instance when the peakednesses (measured by the normalized kurtoses defined as $\text{kurt}_s(n)/E\{s(n)^2\}^2$) and the squared powers of the sources vary inversely proportionally.

The results obtained in [1] and [23] imply in our case that the maxima of the absolute value of $f(v_q, q \in I)$ on the unit sphere are all the points such that only one of the variables $v_q$, with $q \in I$, is nonzero. Equation (9) shows that the output signal $y(n)$ then contains a contribution from only one nonstationary source (and contributions from all stationary sources). Thus, we reach the target partial BSS for one of the nonstationary sources. The last aspect of our method that must be defined is how to select the matrix $\mathbf{B}$ and to constrain the vector $\mathbf{w}$ (which is the parameter controlled in practice, unlike $\mathbf{v}$) so that the variables $\{v_q, q \in I\}$ meet condition (14). First, we define the differential power $\mathrm{Dpow}_y(n_1, n_2)$ of a signal $y$ between the two times $n_1$ and $n_2$ by

$$\mathrm{Dpow}_y(n_1, n_2) = E\{y(n_2)^2\} - E\{y(n_1)^2\}. \qquad (15)$$

By using the independence of the source signals, it may be shown easily that, similarly to the differential kurtosis (12), we have

$$\mathrm{Dpow}_y(n_1, n_2) = \sum_{q \in I} v_q^2 \mathrm{Dpow}_{s_q}(n_1, n_2). \qquad (16)$$

The BSS scale indeterminacy makes it possible to rescale the differential powers of the $P$ nonstationary sources up to positive factors. Therefore, provided these $P$ differential powers are strictly positive for the considered times $n_1$ and $n_2$, they may be assumed to be equal to 1 without loss of generality. Then, (16) reads

$$\mathrm{Dpow}_y(n_1, n_2) = \sum_{q \in I} v_q^2 \qquad (17)$$

so that the constraint (14) can be expressed in terms of unit differential power for the output signal $y(n)$. Then, we introduce a differential extension of the sphering stage of standard kurtotic methods, i.e., we aim at deriving a matrix $\mathbf{B}$ so that the normalization of the differential power of $y(n)$ can be done by normalizing the extraction vector $\mathbf{w}$. To this end, we define the differential correlation matrix of $\mathbf{x}(n)$ as

$$\mathbf{DR}_{\mathbf{x}}(n_1, n_2) = \mathbf{R}_{\mathbf{x}}(n_2) - \mathbf{R}_{\mathbf{x}}(n_1) \qquad (18)$$

[3]In standard approaches, the summation for $q \in I$ in (13) is performed over all $P = N$ sources and the parameters $\alpha_q$ are equal to the standard kurtoses $\mathrm{kurt}_{s_q}(n)$ of all these sources. However, this has no influence on the later discussion, which is based on the general properties of the type of functions defined by (13).

where $\mathbf{R}_{\mathbf{x}}(n_i) = E\{\mathbf{x}(n_i)\mathbf{x}(n_i)^T\}$ is the standard correlation matrix of $\mathbf{x}(n)$ at time $n_i$. Let us now consider the Schur decomposition [26, p. 393] of the real symmetric matrix $\mathbf{DR}_{\mathbf{x}}(n_1, n_2)$

$$\mathbf{DR}_{\mathbf{x}}(n_1, n_2) = \mathbf{E}\boldsymbol{\Delta}\mathbf{E}^T \qquad (19)$$

where $\mathbf{E}$ is a real $P \times P$ orthogonal matrix and $\boldsymbol{\Delta}$ is a $P \times P$ diagonal matrix. It may be shown as follows that the entries of $\boldsymbol{\Delta}$ are strictly positive. Equations (2) and (18) yield

$$\begin{aligned}\mathbf{DR}_{\mathbf{x}}(n_1, n_2) &= \mathbf{R}_{\mathbf{x}}(n_2) - \mathbf{R}_{\mathbf{x}}(n_1) \\ &= \mathbf{AR}_{\mathbf{s}}(n_2)\mathbf{A}^T - \mathbf{AR}_{\mathbf{s}}(n_1)\mathbf{A}^T \\ &= \mathbf{A}\left(\mathbf{R}_{\mathbf{s}}(n_2) - \mathbf{R}_{\mathbf{s}}(n_1)\right)\mathbf{A}^T \\ &= \mathbf{ADR}_{\mathbf{s}}(n_1, n_2)\mathbf{A}^T \end{aligned} \qquad (20)$$

where $\mathbf{DR}_{\mathbf{s}}(n_1, n_2)$ is a diagonal matrix due to source independence and its entries with indices $q \notin I$ are null due to the stationarity of the corresponding sources. It may be checked easily that the columns of $\mathbf{A}$ with indices $q \notin I$ yield no contributions in the right-hand term of (20), so that we also have

$$\mathbf{DR}_{\mathbf{x}}(n_1, n_2) = \check{\mathbf{A}}\boldsymbol{\Lambda}\check{\mathbf{A}}^T \qquad (21)$$

where $\check{\mathbf{A}}$ consists of the $P$ columns of $\mathbf{A}$ with indices $q \in I$ and the diagonal matrix $\boldsymbol{\Lambda}$ contains the differential powers of the $P$ nonstationary sources for times $n_1$ and $n_2$. While (20) involves nonsquare matrices, (21) only contains $P \times P$ matrices. If we assume that $\check{\mathbf{A}}$ is invertible, [26, Th. 4.2.1, p. 141] implies that $\mathbf{DR}_{\mathbf{x}}(n_1, n_2)$ is positive definite if and only if $\boldsymbol{\Lambda}$ is also positive definite. Therefore, if we again assume the differential powers of the nonstationary sources to be strictly positive (for times $n_1$ and $n_2$), (19) and (21) show that all diagonal entries of $\boldsymbol{\Delta}$ are strictly positive. Moreover, if these differential powers are rescaled to unity, then $\boldsymbol{\Lambda} = \mathbf{I}$ and (21) becomes

$$\mathbf{DR}_{\mathbf{x}}(n_1, n_2) = \check{\mathbf{A}}\check{\mathbf{A}}^T. \qquad (22)$$

Since $\boldsymbol{\Delta}$ only has positive entries, it has a real-valued square root, so that we can define $\mathbf{B}$ by

$$\mathbf{B} = \boldsymbol{\Delta}^{-1/2}\mathbf{E}^T. \qquad (23)$$

Then, with $\mathbf{z}$ defined by (3), we have

$$\begin{aligned}\mathbf{DR}_{\mathbf{z}}(n_1, n_2) &= \mathbf{BDR}_{\mathbf{x}}(n_1, n_2)\mathbf{B}^T \\ &= \boldsymbol{\Delta}^{-1/2}\mathbf{E}^T\mathbf{E}\boldsymbol{\Delta}\mathbf{E}^T\mathbf{E}\boldsymbol{\Delta}^{-1/2} \\ &= \mathbf{I}. \end{aligned} \qquad (24)$$

Let us now consider the output signal $y(n)$ defined by (4). We have

$$\begin{aligned}\mathrm{Dpow}_y(n_1, n_2) &= E\{y(n_2)^2\} - E\{y(n_1)^2\} \\ &= \mathbf{w}^T\mathbf{R}_{\mathbf{z}}(n_2)\mathbf{w} - \mathbf{w}^T\mathbf{R}_{\mathbf{z}}(n_1)\mathbf{w} \\ &= \mathbf{w}^T\mathbf{DR}_{\mathbf{z}}(n_1, n_2)\mathbf{w} \\ &= \|\mathbf{w}\|^2. \end{aligned} \qquad (25)$$

By considering the relations (17) and (25), we obtain

$$\|\mathbf{w}\|^2 = \sum_{q \in I} v_q^2. \quad (26)$$

By constraining our vector $\mathbf{w}$ to have unit norm, we hence constrain the vector $\{v_q, q \in I\}$ to be on the unit sphere. Then, as explained previously, the maxima of the absolute value of $\text{Dkurt}_{\mathbf{w}^T\mathbf{z}}(n_1, n_2)$ on the constraint surface $\|\mathbf{w}\| = 1$ rigorously correspond to the partial separation points.

### B. Summary of Proposed Methods

The first practical method which results from the previous analysis consists of the following steps.

Step 1) Select two nonoverlapping bounded time intervals for estimating the statistical parameters (kurtosis, correlation, and power) at two times $n_1$ and $n_2$. In most cases, we obtained better results with intervals such that the differential powers of the sources (which can be roughly evaluated by the differential powers of the observations) are high. These intervals must be such that all nonstationary[4] sources have positive differential powers [defined as in (15)] and nonzero differential kurtoses.[5] The signs of the differential powers of the sources can be obtained by estimating the differential correlation matrix $\mathbf{DR_x}$ and by computing its Schur decomposition $\mathbf{E\Delta E}^T$. Indeed, if all its eigenvalues are strictly positive, we proved that the $P$ sources of interest all have strictly positive differential powers (if all these eigenvalues are strictly negative, we can permute the two time intervals so as to make them all positive).

Step 2) Derive the matrix $\mathbf{B} = \mathbf{\Delta}^{-1/2}\mathbf{E}^T$ from the previously defined matrices $\mathbf{E}$ and $\mathbf{\Delta}$. This matrix performs a "differential sphering" of the observations, i.e., it yields a vector $\mathbf{z}(n)$ defined by (3) which meets (24).

Step 3) Create an output signal $y(n)$ defined by (4), where $\mathbf{w}$ is a vector which satisfies $\|\mathbf{w}\| = 1$ and which is adapted so as to maximize the absolute value of the differential kurtosis of $y(n)$, defined by (6). Various algorithms may be used to achieve this optimization, especially by developing differential versions of algorithms which were previously proposed for the case when $P = N$. The most classical approach is based on gradient ascent [1]. Here, we preferably derive an improved method from the standard fixed-point FastICA algorithm [24], which yields two advantages with respect to the gradient-based approach, i.e., fast convergence and no tunable parameters. Briefly, as the standard FastICA algorithm, our linear instantaneous differential fixed-point ICA algorithm, denoted LI-DFICA hereafter, takes advantage of the fact that if an extraction vector $\mathbf{w}$ optimizes the criterion $|\text{Dkurt}_y(n_1, n_2)|$, then the gradient of $\text{Dkurt}_y(n_1, n_2)$ has the same direction as $\mathbf{w}$ (see [1, p. 178]). Therefore, our update expression uses the gradient of the differential kurtosis which is derived in Appendix A, i.e., it is based on the assignment

$$\mathbf{w} \propto \frac{\partial \text{Dkurt}_y(n_1, n_2)}{\partial \mathbf{w}}. \quad (27)$$

More precisely, starting from a random unit-norm vector $\mathbf{w}$, our LI-DFICA algorithm then consists in iteratively performing the following operations:
1) differential update of $\mathbf{w}$

$$\mathbf{w} \leftarrow E\{\mathbf{z}(n_2)(\mathbf{w}^T\mathbf{z}(n_2))^3\} - E\{\mathbf{z}(n_1)(\mathbf{w}^T\mathbf{z}(n_1))^3\} \\ - 3[\mathbf{R_z}(n_1) + (1 + \mathbf{w}^T\mathbf{R_z}(n_1)\mathbf{w})\mathbf{I}]\mathbf{w} \quad (28)$$

where the statistical parameters are estimated by time averaging;
2) normalization of $\mathbf{w}$, to meet condition $\|\mathbf{w}\| = 1$, i.e.,

$$\mathbf{w} \leftarrow \frac{\mathbf{w}}{\|\mathbf{w}\|}. \quad (29)$$

Step 4) The nonstationary source signal extracted as $y(n)$ in Step 3) is then used to subtract its contributions from all observed signals. The resulting signals are then processed by using again the previously described complete procedure, thus extracting another source, and so on until all nonstationary sources have been extracted. This corresponds to a deflation procedure, as in the standard FastICA method [24], except that a *differential* version of this procedure is required here again. This differential deflation operates in the same way as the standard deflation, except that the statistical parameters are replaced by their differential versions in order to estimate the entries $a_{kl}$ of the mixing matrix $\mathbf{A}$. Indeed, we prove in Appendix B that the scale of the contribution of an extracted source $s_l(n)$ in the $k$th observation can be obtained by estimating the differential correlation $\text{Dcorr}_{yx_k}(n_1, n_2)$ defined by

$$\text{Dcorr}_{yx_k}(n_1, n_2) = E\{y(n_2)x_k(n_2)\} - E\{y(n_1)x_k(n_1)\} \quad (30)$$

which is shown to be equal to the entry $a_{kl}$ of the mixing matrix $\mathbf{A}$.

Instead of the previously described deflation-based version of our LI-DFICA method, a differential extension of the symmetric approach described in [24] can be considered. The resulting symmetric LI-DFICA method consists of the Steps 1) and 2), followed by iterations on the following operations:

---

[4]"Nonstationary" here means "long-term nonstationary." More precisely, all sources should be stationary inside each of two "short" time intervals associated with times $n_1$ and $n_2$, so that their statistics may be estimated for each of these intervals, by time averaging. This corresponds to "short-term stationarity." Then, the aforementioned "sources of interest" (respectively, "noise sources") consist of source signals whose statistics are requested to vary (respectively, not to vary) from one of the considered time intervals to the other one, i.e., sources which are "long-term nonstationary" (respectively, "long-term stationary").

[5]Note that the hypothesis $\text{Dkurt}_{s_q}(n_1, n_2) \neq 0, \forall q \in I$ does not require the source distributions to have different peakednesses at the two times $n_1$ and $n_2$. Indeed, the peakedness is measured by the normalized kurtosis defined as $\text{kurt}_s(n)/E\{s(n)^2\}^2$, so that a source can have different nonnormalized kurtoses but the same peakedness at the two times as soon as this source has different powers at these times. Then, except in the very specific case when the normalized kurtoses and the squared powers of the sources vary inversely proportionally, we are sure that the sources have nonzero differential kurtoses.

1) differential update of $P$ vectors $\mathbf{w}_i$, with $i = 1, \ldots, P$, according to (28);
2) symmetric orthogonalization of the vectors $\mathbf{w}_i$, i.e.,

$$\mathbf{W} \leftarrow (\mathbf{W}\mathbf{W}^T)^{-1/2}\mathbf{W} \quad (31)$$

where $\mathbf{W} = [\mathbf{w}_1, \ldots, \mathbf{w}_P]^T$.

### C. Convergence Proofs

Let us consider the update expressions (28) and (29) of our differential deflation-based algorithm. The output signal $y(n)$ may be expressed with respect to the sources according to (7) with $\mathbf{v} = \mathbf{M}^T\mathbf{w}$ and $\mathbf{M} = \mathbf{BA}$. Moreover, we proved that the differential kurtosis of $y(n)$ only depends on the sources of interest (whose indices are in the set $I$) and on the associated mixing submatrix $\check{\mathbf{A}}$. Therefore, when considering the overall component $\check{y}(n)$ of $y(n)$ corresponding to the nonstationary sources, we can make the change of variable $\check{\mathbf{v}} = \check{\mathbf{M}}^T\mathbf{w}$ where the matrix $\check{\mathbf{M}} = \mathbf{B}\check{\mathbf{A}}$ combines the mixture and the differential whitening processes for the nonstationary sources. Then, we have

$$\check{y}(n) = \check{\mathbf{v}}^T\check{\mathbf{s}}(n) \quad (32)$$

where the column vector $\check{\mathbf{s}}(n)$ only contains the values of the nonstationary sources, i.e., $s_q(n)$ with $q \in I$. For the sake of legibility, we here omit the considered two times $n_1$ and $n_2$ in the notation for differential kurtosis. The update rule

$$\mathbf{w} \leftarrow \frac{\partial \text{Dkurt}_{\mathbf{w}^T\mathbf{z}}}{\partial \mathbf{w}} \quad (33)$$

then reads

$$\mathbf{w} \leftarrow \frac{\partial \check{\mathbf{v}}}{\partial \mathbf{w}} \frac{\partial \text{Dkurt}_{\mathbf{w}^T\mathbf{z}}}{\partial \check{\mathbf{v}}}. \quad (34)$$

Since $\text{Dkurt}_{\mathbf{w}^T\mathbf{z}} = \text{Dkurt}_{\mathbf{v}^T\mathbf{s}} = \text{Dkurt}_{\check{\mathbf{v}}^T\check{\mathbf{s}}}$, we can rewrite (34) as

$$\mathbf{w} \leftarrow \check{\mathbf{M}}\frac{\partial \text{Dkurt}_{\check{\mathbf{v}}^T\check{\mathbf{s}}}}{\partial \check{\mathbf{v}}}. \quad (35)$$

The update expression of $\check{\mathbf{v}}$ then reads

$$\check{\mathbf{v}} = \check{\mathbf{M}}^T\mathbf{w} \leftarrow \check{\mathbf{M}}^T\check{\mathbf{M}}\frac{\partial \text{Dkurt}_{\check{\mathbf{v}}^T\check{\mathbf{s}}}}{\partial \check{\mathbf{v}}}. \quad (36)$$

Combining (3) and (22) yields

$$\begin{aligned}\mathbf{DR}_\mathbf{z}(n_1, n_2) &= \mathbf{B}\mathbf{DR}_\mathbf{x}(n_1, n_2)\mathbf{B}^T \\ &= \mathbf{B}\check{\mathbf{A}}\check{\mathbf{A}}^T\mathbf{B}^T \\ &= \check{\mathbf{M}}\check{\mathbf{M}}^T \end{aligned} \quad (37)$$

and considering (24)

$$\check{\mathbf{M}}\check{\mathbf{M}}^T = \mathbf{I}. \quad (38)$$

$\check{\mathbf{M}}$ is, therefore, orthogonal and we also have

$$\check{\mathbf{M}}^T\check{\mathbf{M}} = \mathbf{I}. \quad (39)$$

Thus, the update expression (36) becomes

$$\check{\mathbf{v}} \leftarrow \frac{\partial \text{Dkurt}_{\check{\mathbf{v}}^T\check{\mathbf{s}}}}{\partial \check{\mathbf{v}}}. \quad (40)$$

By considering (12) and by taking into account that the vector $\check{\mathbf{v}}$ only contains the coefficients of the nonstationary sources $s_i, i \in I$, we can express the update equation of each component of $\check{\mathbf{v}}$ in (40) as

$$\check{v}_i \leftarrow 4\text{Dkurt}_{s_i}\check{v}_i^3 \propto \text{Dkurt}_{s_i}\check{v}_i^3. \quad (41)$$

Choosing $j \in I$ so that $\text{Dkurt}_{s_j} \neq 0$ and $\check{v}_j \neq 0$, we obtain

$$\frac{|\check{v}_i|}{|\check{v}_j|} \leftarrow \frac{|\text{Dkurt}_{s_i}|}{|\text{Dkurt}_{s_j}|}\left(\frac{|\check{v}_i|}{|\check{v}_j|}\right)^3, \quad \forall i \in I. \quad (42)$$

This recursive formula lets us solve analytically $(|\check{v}_i|/|\check{v}_j|)$

$$\frac{|\check{v}_i(k)|}{|\check{v}_j(k)|} = \sqrt{\left|\frac{\text{Dkurt}_{s_j}}{\text{Dkurt}_{s_i}}\right|} \cdot \left(\sqrt{\left|\frac{\text{Dkurt}_{s_i}}{\text{Dkurt}_{s_j}}\right|}\left|\frac{\check{v}_i(0)}{\check{v}_j(0)}\right|\right)^{3^k}. \quad (43)$$

Then, all the components $\check{v}_i(k)$ but $\check{v}_j(k)$ so that $j = \arg\max_p(\sqrt{|\text{Dkurt}_{s_p}|}|\check{v}_p(0)|)$ quickly become small compared to $\check{v}_j(k)$. With the normalization $\|\mathbf{w}\| = 1$ which implies $\|\check{\mathbf{v}}\| = 1$ because of the orthogonality of the matrix $\check{\mathbf{M}}$, we conclude that $\check{v}_j(k) \to \pm 1$ and $\check{v}_i(k) \to 0, \forall i \neq j$. Thus, we proved that $\check{\mathbf{v}}$ converges towards a vector with only one nonzero entry (equal to $\pm 1$). This yields the partial separation of the sources of interest, as $\check{y}(n) = \check{\mathbf{v}}^T\check{\mathbf{s}}(n)$. Thus, we proved the global convergence [i.e., whatever $\check{\mathbf{v}}(0)$] of our algorithm. Moreover, this convergence is cubic (as with the (over)determined FastICA algorithm [24]), which means very fast convergence.

For our symmetric algorithm, the same approach as in [27] can be used in the differential case. Indeed, in the (over)determined case, Oja proved that when the updates obey the rules

$$v_{pq} \leftarrow \text{kurt}_{s_q}v_{pq}^3 \quad (44)$$

for the $q$th component of each extraction vector $\mathbf{v}_p$, these updates combined with the symmetric orthogonalization stage $\mathbf{W} \leftarrow (\mathbf{W}\mathbf{W}^T)^{-1/2}\mathbf{W}$ imply cubic convergence to the separation points and the stability of these separation points, as in the deflation algorithm. By noticing that our differential algorithm yields (41), which is equivalent to (44) with $q = i$, except that the standard kurtoses $\text{kurt}_{s_q}$ are replaced by their differential versions $\text{Dkurt}_{s_i}$, we thus prove rigorously the cubic convergence of our symmetric differential algorithm.

## III. PROPOSED DIFFERENTIAL BSS METHOD FOR CONVOLUTIVE MIXTURES

### A. Mixture Model and Goal

Here, we consider convolutive mixtures defined by a set of $P \times N$ unknown filters with impulse responses $a_{ij}(n)$ which form a matrix function $\mathbf{A}(n) = [a_{ij}(n)]$. The overall relationship between the source and observation vectors $\mathbf{s}(n)$ and $\mathbf{x}(n)$ then reads

$$\mathbf{x}(n) = \mathbf{A}(n) * \mathbf{s}(n). \quad (45)$$

Each source $s_j(n)$ is here assumed to be expressed as

$$s_j(n) = f_j(n) * u_j(n) \quad (46)$$

where $f_j(n)$ is a filter impulse response and $u_j(n)$ is the innovation process of $s_j(n)$. Denoting $\mathbf{u}(n) = [u_1(n), \ldots, u_N(n)]^T$, we can then express the mixing (45) as

$$\mathbf{x}(n) = \mathbf{H}(n) * \mathbf{u}(n) \quad (47)$$

where $\mathbf{H}(n) = \mathbf{A}(n) * \mathbf{F}(n)$ with $\mathbf{F}(n) = \mathrm{diag}(f_1(n), \ldots, f_N(n))$.

We make the following assumptions concerning the aforementioned mixture model.

- The process $\mathbf{u}(n) = [u_1(n), \ldots, u_N(n)]^T$ is real-valued, zero-mean, temporally independent and identically distributed (i.i.d.) and spatially independent, i.e., its components $u_j(n)$ are statistically independent from each other and do not necessarily have the same distribution.
- The function matrices $\mathbf{A}(n), \mathbf{F}(n)$ and thus $\mathbf{H}(n)$ correspond to causal and finite impulse response (FIR) filters and are nonsingular. Note that infinite impulse response (IIR) systems can also be approximated by equivalent (high-order) FIR models.

Convolutive BSS typically aims at estimating the contributions of all sources in each observation, i.e., $a_{ij}(n) * s_j(n)$. In convolutive deflation-based methods such as [28], this is achieved as follows.

1) Extract the innovation process $u_j(n)$ of a source $s_j(n)$ from the observations.
2) Identify $P$ coloring filters and apply them to $u_j(n)$ in order to recover the contributions of $s_j(n)$ in each observation.
3) Subtract these contributions from all the observations.
4) Set $N \leftarrow N - 1$. If $N \neq 1$, go back to Step 1) in order to extract another source.

### B. Previously Reported Approach for (Over)Determined Convolutive Mixtures

Recently, we proposed a time-domain fast fixed-point algorithm for determined (i.e., $P = N$) and overdetermined (i.e., $P > N$) convolutive mixtures [25]. This algorithm, denoted C-FICA hereafter, may be seen as a convolutive extension of FastICA and keeps its attractive features. Here, we briefly describe the principles of the kurtotic version of this previous approach, as we will then apply our differential BSS concept to that algorithm in order to introduce a new partial separation method for underdetermined mixtures involving some nonstationary sources. The first step of our (over)determined approach performs a "convolutive sphering" of the observations, defined as follows. At any time $n$, we consider the column vector

$$\tilde{\mathbf{x}}(n) = [x_1(n+R), \ldots, x_1(n-R), \ldots,$$
$$x_P(n+R), \ldots, x_P(n-R)]^T \quad (48)$$

which contains $M = (2R+1)P$ entries. We derive the $M$-entry column vector $\tilde{\mathbf{z}}(n) = [\tilde{z}_1(n), \ldots, \tilde{z}_M(n)]^T$ defined as

$$\tilde{\mathbf{z}}(n) = \tilde{\mathbf{B}} \tilde{\mathbf{x}}(n) \quad (49)$$

where $\tilde{\mathbf{B}}$ is an $M \times M$ matrix chosen so that

$$\mathbf{R}_{\tilde{\mathbf{z}}} = E\{\tilde{\mathbf{z}} \tilde{\mathbf{z}}^T\} = \mathbf{I}. \quad (50)$$

With respect to $\tilde{\mathbf{x}}(n)$, operation (49) may, therefore, be considered as conventional sphering, which consists of principal component analysis and normalization. Now, with respect to the original observations $x_i(n)$, this may be interpreted differently: Equations (48) and (49) show that the signals $\tilde{z}_i(n)$ are convolutive mixtures of the signals $x_i(n)$. Equation (50) then means that the signals $\tilde{z}_i(n)$ are created so as to have unit variances and to be mutually uncorrelated, which may be seen as a spatio–temporal whitening and normalization of the observations $x_i(n)$.

Let us denote by $y(n)$ the extracted signal

$$y(n) = \mathbf{w}^T \tilde{\mathbf{z}}(n) = \sum_{m=1}^{M} w_m \tilde{z}_m(n) \quad (51)$$

where $\mathbf{w}$ is an $M$-entry extended column vector of extraction coefficients $w_m$ which, together with (48) and (49), yields a convolutive combination $y(n)$ of the observations. The power of $y(n)$ reads $E\{y(n)^2\} = \mathbf{w}^T \mathbf{R}_{\tilde{\mathbf{z}}} \mathbf{w}$. By constraining $\tilde{\mathbf{z}}(n)$ so as to meet (50), we get $E\{y(n)^2\} = \mathbf{w}^T \mathbf{I} \mathbf{w} = \|\mathbf{w}\|^2$. Our method then consists in maximizing the absolute value of the nonnormalized kurtosis of $y(n)$ defined by (51) under the constraint $\|\mathbf{w}\| = 1$ so that $y(n)$ has unit power. We proved in [25] that this constrained optimization yields an estimate $e_l(n)$ of a delayed and scaled source innovation process $\beta_l u_l(n - r_l)$, under some conditions. Therefore, this C-FICA algorithm based on our modified vector $\mathbf{w}$ extracts one source innovation process as follows.

- Initialize $\mathbf{w}$ to a value $\mathbf{w}_0$, e.g., using the approaches presented below.
- Repeat the following Steps 1) and 2) until convergence

  1) $\mathbf{w} \leftarrow E\{\tilde{\mathbf{z}}(\mathbf{w}^T \tilde{\mathbf{z}})^3\} - 3\mathbf{w} \propto \dfrac{\partial \mathrm{kurt}_{\mathbf{w}^T \tilde{\mathbf{z}}}(n)}{\partial \mathbf{w}}$ (52)
  
  2) $\mathbf{w} \leftarrow \dfrac{\mathbf{w}}{\|\mathbf{w}\|}.$ (53)

The aforementioned initial value $\mathbf{w}_0$ of $\mathbf{w}$ may be selected randomly. An improved approach was derived in [25] from the relationship which exists between our vector $\mathbf{w}$ and the coefficients of the FIR filters $k_p(n)$ of Tugnait's approach [28], which derives an output signal $y(n)$ as

$$y(n) = \sum_{p=1}^{P} k_p(n) * x_p(n) = \sum_{p=1}^{P} \sum_{r=-R}^{R} k_p(r) x_p(n-r) \quad (54)$$

where $k_p(n), p = 1 \ldots P$ are $P$ noncausal FIR filters in practice. Indeed, in [25], we proved that

$$\tilde{\mathbf{k}} = \mathbf{w}^T \tilde{\mathbf{B}} \quad (55)$$

where the row vector $\tilde{\mathbf{k}}$ consists of the impulse response coefficients of the filters $k_1(n)$ to $k_P(n)$. This relation lets us initialize our vector $\mathbf{w}$ as in Tugnait's method, i.e., with unit filters $k_p(n) = \delta(n)$, so that $y(n)$ is the sum of all observations $x_p(n)$. That corresponds to $\tilde{\mathbf{k}} = \tilde{\mathbf{k}}_0$ defined as

$$\tilde{\mathbf{k}}_0 = [\underbrace{0, \ldots, 0, 1, 0, \ldots, 0}_{k_1}, \ldots \ldots, \underbrace{0, \ldots, 0, 1, 0, \ldots, 0}_{k_P}]. \quad (56)$$

Equation (55) then yields

$$\mathbf{w}_0^T = \tilde{\mathbf{k}}_0 \tilde{\mathbf{B}}^{-1}. \quad (57)$$

This initialization of $\mathbf{w}$ provided better experimental results than a random one and is also used in Section III-C for underdetermined convolutive mixtures.

As stated previously, this extraction stage provides an estimate $e_l(n)$ of a source innovation process up to a delay and a scale factor. Then, we can color it to obtain each contribution of the $l$th source in the $k$th observation $x_k(n)$. This can be done by deriving the noncausal coloration filters $C_{kl}(z) = \sum_{r=-R'}^{R'} c_{kl}(r) z^{-r}$ which make the signals $c_{kl}(n) * e_l(n)$ be the closest to $x_k(n)$ in the mean square sense [29]. This was achieved in [25] by noncausal FIR Wiener filters [30], whose impulse response coefficients form vectors $\mathbf{c}_{kl}$ defined by

$$\mathbf{c}_{kl} = \mathbf{R}_{e_l}^{-1} \mathbf{r}_{e_l x_k} \quad (58)$$

where $\mathbf{R}_{e_l}$ is the autocorrelation matrix of the signal $e_l(n)$ and $\mathbf{r}_{e_l x_k}$ is the cross-correlation vector of the signals $e_l(n)$ and $x_k(n)$. Note that the autocorrelation matrix has a highly regular Toeplitz structure and there are a number of efficient methods [30] for solving the linear matrix equation (58). After subtracting the contributions $c_{kl}(n) * e_l(n)$ from all observations, we obtain another mixture configuration with $N - 1$ sources. The extraction stage must then be iterated as explained in Section III-A to extract the innovation process of another source.

### C. Extension of Convolutive Approach to Underdetermined Mixtures

Here, we aim at extending our C-FICA algorithm for convolutive BSS to the underdetermined case, using the differential BSS concept that we introduced in [22]. The resulting algorithm is therefore denoted C-DFICA hereafter. Let us again define $y(n)$ by (48), (49), and (51) where $\tilde{\mathbf{B}}$ and $\mathbf{w}$ will be selected as explained further in this section. As in Section II, we denote by $\mathrm{kurt}_y(n_1)$ and $\mathrm{kurt}_y(n_2)$, respectively, the nonnormalized kurtoses of $y(n)$ for two times $n_1$ and $n_2$, and we define its nonnormalized differential kurtosis by (6). Combining the definition (48), (49), and (51) of $y(n)$ with (47) shows that $y(n)$ is a linear combination of delayed versions of the processes $u_k(n)$ i.e.,

$$y(n) = \sum_{q=1}^{N} \sum_{d=D_{\min}}^{D_{\max}} v_{qd} u_q(n - d) \quad (59)$$

where $D_{\min}$ and $D_{\max}$ are derived from the orders of the FIR filters involved in (47) and (48). The processes $u_q(n)$ are here still assumed to be real-valued, zero-mean, and mutually and temporally independent. $P$ of them are now assumed to be long-term nonstationary (i.e., not identically distributed) and correspond to the sources of interest, while the other $N - P$ are stationary and correspond to the "noise sources." Using the multilinearity of the kurtosis for independent random variables, we derive from (6) and (59)

$$\mathrm{Dkurt}_y(n_1, n_2) = \sum_{q=1}^{N} \sum_{d=D_{\min}}^{D_{\max}} v_{qd}^4 \left[ \mathrm{kurt}_{u_q}(n_2 - d) - \mathrm{kurt}_{u_q}(n_1 - d) \right]$$

$$= \sum_{q=1}^{N} \sum_{d=D_{\min}}^{D_{\max}} v_{qd}^4 \alpha_{qd} \quad (60)$$

where

$$\alpha_{qd} = \mathrm{Dkurt}_{u_q}(n_1 - d, n_2 - d) \quad (61)$$

is defined as in (6). As in the linear instantaneous case, let us denote by $I$ the set containing the $P$ indices of the unknown nonstationary sources. The previous hypotheses of our differential concept mean that only the processes $u_q(n)$ with $q \notin I$ are assumed to have the same kurtosis for different times, i.e.,

$$\forall q \notin I \quad \forall d \in [D_{\min} \ldots D_{\max}]$$
$$\mathrm{kurt}_{u_q}(n_1 - d) = \mathrm{kurt}_{u_q}(n_2 - d). \quad (62)$$

Thus, we have

$$\mathrm{Dkurt}_y(n_1, n_2) = \sum_{q \in I} \sum_{d=D_{\min}}^{D_{\max}} v_{qd}^4 \alpha_{qd}. \quad (63)$$

Hence, we see that the noise sources (whose indices do not belong to the set $I$) are invisible in our differential extraction criterion. As an extension of the linear instantaneous case, we here consider the values of (63) on the $P \times D$ dimensional unit sphere[6] where $D = D_{\max} - D_{\min} + 1$, i.e., $\{v_{qd}, q \in I, d \in [D_{\min}, \ldots, D_{\max}]\}$ such that

$$\sum_{q \in I} \sum_{d=D_{\min}}^{D_{\max}} v_{qd}^2 = 1. \quad (64)$$

First, it may be shown easily that the differential power of $y(n)$, again defined by (15), here reads

$$\mathrm{Dpow}_y(n_1, n_2) = \sum_{q \in I} \sum_{d=D_{\min}}^{D_{\max}} v_{qd}^2 \mathrm{Dpow}_{u_q}(n_1 - d, n_2 - d). \quad (65)$$

Each process $u_q(n)$ with $q \in I$ is assumed to exhibit nonstationarity from time $n_1$ to $n_2$. However, it is here assumed to be identically distributed for all times $n_i - d$ when $d$ is varied from $D_{\min}$ to $D_{\max}$. In practice, this means that we select $n_1$ and $n_2$ such that $n_2 - n_1 \gg D$, and that $u_q(n)$ should exhibit long-term nonstationarity between the times $n_1$ and $n_2$ but short-term stationarity inside each time window $[n_i - D_{\min}, n_i - D_{\max}]$ with $i = 1$ or 2. Then, the terms $\mathrm{Dpow}_{u_q}(n_1 - d, n_2 - d)$ in (65) do not depend on $d$, i.e., (65) only involves a single differential power $\mathrm{Dpow}_{u_q}(n_1, n_2)$ for each nonstationary source. These differential powers may again be rescaled by positive factors because of the BSS scale ambiguity. Hence, when they are strictly positive, they may be assumed to be equal to 1 so that (65) becomes

$$\mathrm{Dpow}_y(n_1, n_2) = \sum_{q \in I} \sum_{d=D_{\min}}^{D_{\max}} v_{qd}^2. \quad (66)$$

Then, if $\mathbf{DR}_{\tilde{\mathbf{x}}}(n_1, n_2) = \mathbf{E} \Delta \mathbf{E}^T$ is the symmetric Schur decomposition of $\mathbf{DR}_{\tilde{\mathbf{x}}}(n_1, n_2)$ defined for $\tilde{\mathbf{x}}$ as in (18), we define $\tilde{\mathbf{B}}$ by $\tilde{\mathbf{B}} = \Delta^{-1/2} \mathbf{E}^T$. Then, since we defined $\tilde{\mathbf{z}}$ by (49), we can prove as in (24) that

$$\mathbf{DR}_{\tilde{\mathbf{z}}}(n_1, n_2) = \mathbf{I} \quad (67)$$

which implies, using the same approach as in (25) but now with (51)

$$\mathrm{Dpow}_y(n_1, n_2) = \|\mathbf{w}\|^2 \quad (68)$$

[6]Convolutive mixtures entail a slight approximation concerning which points of this sphere may be reached, as explained in Section III-E.

so that due to (66)

$$\|\mathbf{w}\|^2 = \sum_{q \in I} \sum_{d=D_{\min}}^{D_{\max}} v_{qd}^2. \tag{69}$$

Therefore, varying $\mathbf{w}$ so that $\|\mathbf{w}\| = 1$ yields a simple method in the underdetermined convolutive case to constrain our vector $\{v_{qd}, q \in I, d \in [D_{\min}, \ldots, D_{\max}]\}$ to be on the unit sphere (64). The optimization of the absolute value of $\mathrm{Dkurt}_y(n_1, n_2)$ defined in (63) under the constraint (64), therefore, belongs to the same generic problem as in (13)–(14), with a set of variables here denoted $v_{qd}$ instead of $v_q$. Applying again the results of [1] and [23] that are used in Section II-A here guarantees that the maxima of $|\mathrm{Dkurt}_y(n_1, n_2)|$ on the unit sphere correspond to all the points such that only one of the considered variables $v_{qd}$, with $q \in I$, is nonzero. Equation (59) shows that the output signal then only consists of a delayed and scaled version $e_l(n)$ of the process $u_l(n)$ associated to a nonstationary source (and contributions from all stationary sources). We then color $e_l(n)$ to estimate its contributions in all observations. This is done by adapting the Wiener solution that we used in [25], i.e., by introducing a differential Wiener filtering process, that we define in Section III-D.

### D. Summary of Proposed Method

We propose the following procedure to achieve the fast partial separation of underdetermined convolutive mixtures. This procedure is based on an adapted fixed-point algorithm and on our differential Wiener filtering process in order to recover the contributions of the sources in the observations in the framework of a deflation approach.

Step 1) Select two nonoverlapping bounded time intervals using the same type of approach as in Section II-B.

Step 2) Compute an estimate of the differential correlation matrix $\mathbf{DR}_{\tilde{\mathbf{x}}}(n_1, n_2)$ of the expanded observation vector $\tilde{\mathbf{x}}$. Then, perform the symmetric Schur decomposition of that matrix. This yields a matrix $\mathbf{E}$ whose columns are the unit-norm eigenvectors of the estimate of $\mathbf{DR}_{\tilde{\mathbf{x}}}(n_1, n_2)$ and a diagonal matrix $\mathbf{\Delta}$ which contains the eigenvalues of the estimate of $\mathbf{DR}_{\tilde{\mathbf{x}}}(n_1, n_2)$. Then, derive the matrix $\tilde{\mathbf{B}} = \mathbf{\Delta}^{-1/2} \mathbf{E}^T$. We thus obtain a vector $\tilde{\mathbf{z}}(n)$ defined by (49) which meets (67).

Step 3) Initialize the extraction vector $\mathbf{w}$, using (57) as in the C-FICA algorithm but with the new matrix $\tilde{\mathbf{B}}$ as defined in Step 2. Create an output signal $y(n)$ defined by (51), where $\mathbf{w}$ is a normalized vector which is adapted so as to maximize the absolute value of the differential kurtosis of $y(n)$. The same approach as in Step 3) of Section II-B may be used to this end. This here yields the following C-DFICA algorithm:

a) differential update of $\mathbf{w}$

$$\begin{aligned}\mathbf{w} &\leftarrow E\{\tilde{\mathbf{z}}(n_2)(\mathbf{w}^T\tilde{\mathbf{z}}(n_2))^3\} - E\{\tilde{\mathbf{z}}(n_1)(\mathbf{w}^T\tilde{\mathbf{z}}(n_1))^3\} \\ &\quad - 3[\mathbf{R}_{\tilde{\mathbf{z}}}(n_1) + (1 + \mathbf{w}^T\mathbf{R}_{\tilde{\mathbf{z}}}(n_1)\mathbf{w})\mathbf{I}]\mathbf{w} \\ &\propto \frac{\partial \mathrm{Dkurt}_y(n_1, n_2)}{\partial \mathbf{w}};\end{aligned} \tag{70}$$

b) normalization of $\mathbf{w}$, to meet condition $\|\mathbf{w}\| = 1$, i.e.,

$$\mathbf{w} \leftarrow \frac{\mathbf{w}}{\|\mathbf{w}\|}. \tag{71}$$

Step 4) The nonstationary process $u_l(n)$ extracted as $y(n)$ in Step 3) is then colored by means of our differential Wiener filtering process so as to recover the contributions of the corresponding source in all observations. This filter minimizes the differential power of the difference between an observation $x_k(n)$ and a noncausal filtered version $c_{kl}(n) * y(n)$ of the signal $y(n)$. In Appendix C, we show the consistency of this minimization and we derive the expression of the coloration filters which achieve it, i.e.,

$$\mathbf{c}_{kl} = (\mathbf{R}_{e_l}(n_2) - \mathbf{R}_{e_l}(n_1))^{-1}(\mathbf{r}_{e_l x_k}(n_2) - \mathbf{r}_{e_l x_k}(n_1)). \tag{72}$$

As with the kurtotic and power statistical functions, Appendix C proves that this coloration process does not depend on the noise sources. The resulting signals $c_{kl}(n) * y(n)$ are then subtracted from the observations. Steps 2)–4) are then iterated until we have extracted all source innovation processes.

We do not introduce here a symmetric version for this differential algorithm intended for convolutive mixtures. Indeed, as detailed in Section III-E, the considered convolutive mixtures may also be expressed in terms of instantaneous mixtures and therefore correspond to a considerably higher dimensionality than in Section II-B for our optimization space. Then, we save major computational and memory resources by extracting the innovation processes one by one.

### E. Convergence Proof

Inside each of the mixed signals of $\tilde{\mathbf{x}}(n)$ defined by (48), we here only consider the overall component which corresponds to the nonstationary sources, since the above BSS criterion only depends on this component. The analysis in [25] shows that these convolutive mixtures can be interpreted as linear instantaneous mixtures of multiple delayed and scaled versions of the processes $u_q(n)$ with $q \in I$. In particular, that analysis entails that if $Q$ denotes the order of $\mathbf{G}(n)$ in (47), and if

$$PL \geq \check{N}(Q + L) \tag{73}$$

is met, where $L = (2R + 1)$ is the number of lags, $P$ is the number of mixtures and $\check{N}$ is the number of nonstationary sources, then $\tilde{\mathbf{x}}(n)$ is an *(over)determined* instantaneous mixture of delayed and scaled versions of the processes $u_q(n)$ with $q \in I$. The convergence proof obtained in Section II-C in the instantaneous case can then be applied to our differential convolutive algorithm and, therefore, guarantees global cubic convergence here again. If (73) is not met, which is especially the case when $P = \check{N}$, the *reformulated instantaneous* mixture is underdetermined. This underdetermination is related to the finite order of the equivalent extraction filters in (54). However, when the ratio $PL/[\check{N}(Q + L)]$ associated with (73) tends to 1 (which is the case when $P = \check{N}$ and $L$ is large), our configuration is nearly determined i.e., the equivalent instantaneous mixing matrix is almost square. Then, one still almost obtains an estimate of one process $u_q(n)$ by maximizing the absolute value of the differential nonnormalized kurtosis under unit differential power constraint.[7]

## IV. Experimental Results

Here, we aim at comparing the performance of our new methods for underdetermined mixtures with the nondifferential ones i.e., the FastICA and C-FICA algorithms. First, we tested our deflation method LI-DFICA for linear instantaneous mixtures in the case $N = 5$ and $P = \check{N} = 2$, where the two nonstationary sources correspond to some bass and piano sound signals [31] with 100 000 samples on each of the time domains $D_1$ and $D_2$. The three noise sources are stationary signals corresponding to uniform, Gaussian, and Laplacian distributions with the same power. Their scale factors and resulting powers are varied in our tests in order to investigate the evolution of performance with respect to the input signal-to-noise ratio (SNR$_\text{in}$) of the observations. Our SNR$_\text{in}$ criterion reads

$$\text{SNR}_\text{in} = \text{mean}_i \, 10 \log_{10} \left( \frac{\sum_{j \in I} E\{x_{ij}^2\}}{\sum_{j \notin I} E\{x_{ij}^2\}} \right) \quad (74)$$

where $x_{ij}(n)$ is the contribution of the $j$th source on the $i$th sensor. We average the two SNR$_\text{in}$ values associated with the two used time domains.

For each scale factor applied to the normalized noise sources, we made 100 Monte Carlo simulations by varying the coefficients of the mixing matrix with a uniform distribution in $[-0.5, 0.5]$. The performance of our BSS method is measured by its output signal-to-interference ratio (SIR$_\text{out}$), which may be compared to the input SIR (SIR$_\text{in}$) associated to the observations. The definitions of these parameters are provided in Appendix D.

Fig. 1(a) shows that whatever the SNR$_\text{in}$, the SIR$_\text{out}$ is higher for our differential algorithm compared to the standard one. Moreover, performance begins to fall significantly for SNR$_\text{in}$ smaller than 10 dB for our differential algorithm, instead of 35 dB for the standard one. Besides, for our differential version, the SIR$_\text{out}$ is greater than 30 dB for the first extracted source for SNR$_\text{in}$ down to about 0 dB, instead of 20 dB for the standard algorithm. These results are obtained for a mean SIR$_\text{in}$ of 5 dB.

We also compared the symmetric version of our instantaneous differential algorithm with the standard symmetric FasICA algorithm in the same mixture configuration as previously. Another criterion is commonly used to evaluate performance for symmetric algorithms: We can directly consider the estimated nonstationary sources and compare them with the normalized original sources. This can be done by using the $P \times N$ performance matrix $\mathbf{G} = [g_{ij}]$ which is the product of the mixing matrix and the estimated partial separation matrix. The performance index for the $j$th source signal then reads

$$\forall j \in I, \quad \text{Perf}(j) = \max_i 10 \log_{10} \left( \frac{g_{ij}^2}{\sum_{k \in I}^{k \neq j} g_{ik}^2} \right). \quad (75)$$

---

[7]The approximation that we mentioned in footnote 6 concerning the points of the unit sphere which may be reached in the convolutive case is related as follows to the previous discussion. In the differential linear instantaneous case, $\check{\mathbf{v}} = \check{\mathbf{M}}^T \mathbf{w}$, where $\check{\mathbf{M}}$ is an orthogonal matrix. Therefore, varying $\mathbf{w}$ makes it possible to reach any value of $\check{\mathbf{v}}$, especially including all the unit sphere. The same result applies to convolutive mixtures if (73) is met because, as explained previously, these mixtures can then be reformulated as linear instantaneous ones. On the contrary, when considering convolutive mixtures such that (73) is not met, one cannot guarantee that any point $\{v_{qd}, q \in I, d \in [D_\text{min}, \ldots, D_\text{max}]\}$ on the unit sphere can be reached by varying $\mathbf{w}$. Still, this effect becomes negligible when $PL/[\check{N}(K+L)]$ gets close to 1, as explained previously. This phenomenon is, therefore, ignored in this paper.

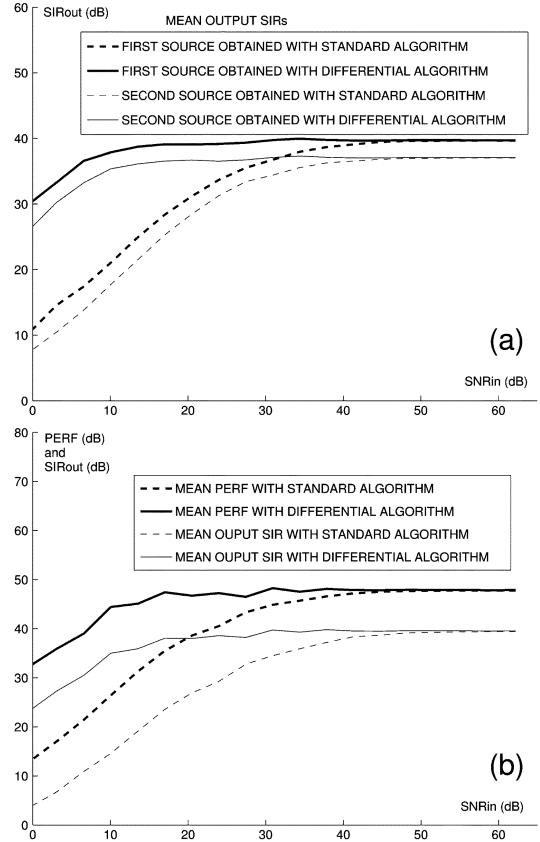

Fig. 1. Output SIR of FastICA and LI-DFICA depending on the input SNR: (a) deflation algorithm and (b) symmetric algorithm.

The resulting $\text{Perf}(j)$ is averaged with respect to the index $j$ so as to obtain a unique criterion Perf for a given performance matrix $\mathbf{G}$.

Fig. 1(b) shows the same advantages as previously described for our differential algorithm compared to the standard one. This time, the performance index Perf begins to fall significantly for SNR$_\text{in}$ smaller than 15 dB (respectively, 35 dB) for the differential algorithm (respectively, the standard algorithm) and this criterion stays higher than 33 dB for SNR$_\text{in}$ down to 0 dB (respectively, 16 dB). In this figure, the SIR$_\text{out}$ of the source contributions, estimated by the differential correlation (30) as a postprocessing stage, is also represented. This shows for these symmetric algorithms that the Perf criterion involving power normalization yields higher performance figures than the SIR$_\text{out}$ criterion, which requires source contribution estimation. Note that these SIR$_\text{out}$ of source contributions are slightly greater than those obtained with the deflation algorithm. In the convolutive case, we tested our C-FICA and C-DFICA algorithms for two long-term nonstationary artificial tenth-order colored sources driven by Laplacian processes. They are short-term stationary on two time windows of 100 000 samples. We used real mixing filters whose sixty fourth-order impulse responses were measured at the ears of a dummy head

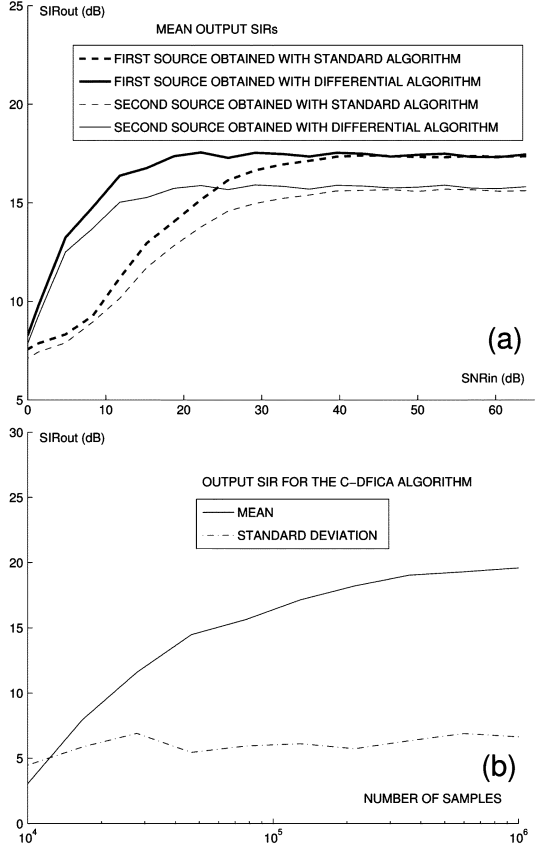

Fig. 2. Output SIR: (a) C-FICA and C-DFICA depending on the input SNR and (b) C-DFICA depending on the number of samples.

[32]. As in the linear instantaneous case, the noise sources are scaled uniform, Gaussian, and Laplacian signals whose scale factors are varied in order to change the $\text{SNR}_{\text{in}}$ defined by (74). For each scale factor applied to the noise sources, we considered 153 different filter sets associated with different source positions. We only constrained the two sources of interest not to be very close to each other (the angular difference was taken greater than or equal to $20°$) and placed the noise sources at positions associated with the angles $-90°$, $30°$, and $150°$. As with the deflation versions of the instantaneous algorithms, we express performance in terms of $\text{SIR}_{\text{out}}$ of source contributions (measured as explained in Appendix D), this time obtained by using the differential Wiener process (72).

Fig. 2(a) represents the $\text{SIR}_{\text{out}}$ of the standard and differential fixed-point ICA algorithms depending on the input SNR. As in the linear instantaneous case, the performance of our differential algorithm stays higher than the C-FICA algorithm whatever $\text{SNR}_{\text{in}}$. More precisely, it begins to fall significantly for $\text{SNR}_{\text{in}}$ smaller than 15 dB (respectively, 35 dB) for our differential version (respectively, the standard one). For the first extracted source, the $\text{SIR}_{\text{out}}$ stays greater than 15 dB for $\text{SNR}_{\text{in}}$ down to 8 dB (respectively, 22 dB). In the last experiment, we aim at evaluating the robustness of our differential algorithm for convolutive mixtures depending on the number of samples available in each time domain. To this end, we fixed the scale factor applied to the stationary noises so as to obtain a mean $\text{SNR}_{\text{in}}$ of 11 dB over all source positions and for each number of samples $n$, we tested our differential algorithm on the 153 filter sets used in the previous experiment. Fig. 2(b) shows that the mean $\text{SIR}_{\text{out}}$ stays higher than 15 dB for $n$ down to 50 000 samples.

## V. CONCLUSION

In this paper, we have introduced new fast fixed-point algorithms for BSS in the case of linear instantaneous and convolutive underdetermined mixtures. They are inspired from the well-known FastICA algorithm and from our recent C-FICA method only intended for (over)determined convolutive mixtures. The LI-DFICA and C-DFICA approaches proposed here are based on a new criterion, called differential nonnormalized kurtosis, that exploits the stationarity of some "noise" sources to achieve the partial separation of the sources of interest. We introduced differential sphering processes so as to use parameter-free fast fixed-point algorithms to achieve the optimization of this new criterion and differential correlation and Wiener filtering methods to recover the source contributions in each observation. As with the FastICA algorithm, a symmetric version of LI-DFICA is also proposed in the instantaneous case. In addition to the attractivity of the limited restrictions that we set about the sources compared to some other approaches for underdetermined mixtures, experimental results show the efficiency of our algorithms in comparison with the associated (over)determined FastICA and C-FICA algorithms.

## APPENDIX A
## GRADIENT OF THE DIFFERENTIAL KURTOSIS

The nonnormalized kurtosis of a zero-mean signal $y(n) = \mathbf{w}^T \mathbf{z}(n)$ is defined by

$$\text{kurt}_y(n) = E\{y^4(n)\} - 3[E\{y^2(n)\}]^2. \quad (76)$$

By taking its first-order derivative with respect to $\mathbf{w}$, we obtain

$$\begin{aligned}\frac{\partial \text{kurt}_y(n)}{\partial \mathbf{w}} &= \frac{\partial}{\partial \mathbf{w}}(E\{(\mathbf{w}^T\mathbf{z}(n))^4\} - 3[E\{(\mathbf{w}^T\mathbf{z}(n))^2\}]^2)\\ &= E\{4\mathbf{z}(n)(\mathbf{w}^T\mathbf{z}(n))^3\} - 6E\{(\mathbf{w}^T\mathbf{z}(n))^2\}E\{2\mathbf{z}(n)\mathbf{w}^T\mathbf{z}(n)\}\\ &= 4E\{\mathbf{z}(n)y^3(n)\} - 12E\{y^2\}E\{\mathbf{z}(n)\mathbf{z}(n)^T\}\mathbf{w}\\ &\propto E\{\mathbf{z}(n)y^3(n)\} - 3E\{y^2(n)\}\mathbf{R}_{\mathbf{z}(n)}(n)\mathbf{w}. \end{aligned} \quad (77)$$

Note that in the nondifferential FastICA approach, the gradient expression (77) can be simplified because the sphering process implies that $E\{y^2(n)\} = 1$ and $\mathbf{R}_{\mathbf{z}}(n) = \mathbf{I}$. The gradient expression is thus

$$\frac{\partial \text{kurt}_y(n)}{\partial \mathbf{w}} \propto E\{\mathbf{z}(n)y^3(n)\} - 3\mathbf{w}. \quad (78)$$

In our approach, the differential gradient reads

$$\begin{aligned}\frac{\partial \text{Dkurt}_y(n_1, n_2)}{\partial \mathbf{w}} &= \frac{\partial \text{kurt}_y(n_2)}{\partial \mathbf{w}} - \frac{\partial \text{kurt}_y(n_1)}{\partial \mathbf{w}}\\ &\propto E\{\mathbf{z}(n_2)y^3(n_2)\} - 3E\{y^2(n_2)\}\mathbf{R}_{\mathbf{z}}(n_2)\mathbf{w}\\ &\quad - E\{\mathbf{z}(n_1)y^3(n_1)\} + 3E\{y^2(n_1)\}\mathbf{R}_{\mathbf{z}}(n_1)\mathbf{w}. \end{aligned} \quad (79)$$

Using the relations $\mathbf{R_z}(n_2) - \mathbf{R_z}(n_1) = \mathbf{I}, E\{y^2(n_i)\} = \mathbf{w}^T\mathbf{R_z}(n_i)\mathbf{w} \; \forall i \in \{1,2\}$ and $E\{y^2(n_2)\} - E\{y^2(n_1)\} = 1$, we can simplify expression (79) to obtain

$$\frac{\partial \mathrm{Dkurt}_y(n_1, n_2)}{\partial \mathbf{w}} \propto E\{\mathbf{z}(n_2)y^3(n_2)\} - E\{\mathbf{z}(n_1)y^3(n_1)\} \\ - 3[\mathbf{R_z}(n_1) + (1 + \mathbf{w}^T\mathbf{R_z}(n_1)\mathbf{w})\mathbf{I}]\mathbf{w}. \quad (80)$$

## APPENDIX B
## DIFFERENTIAL CORRELATION

Here, we suppose that the extracted signal only contains one source of interest superimposed with noise sources so that

$$y(n) = \varepsilon s_l(n) + \sum_{q \notin I} \alpha_q s_q(n) \quad (81)$$

with $l \in I$. The scale factor $\varepsilon$ associated with $s_l(n)$ in (81) can be deduced from the differential power of $y(n)$ constrained to be equal to 1 and from the unit differential powers of the nonstationary sources, i.e.,

$$\mathrm{Dpow}_{s_l}(n_1, n_2) = 1 \qquad \forall l \in I \quad (82)$$

which imply that $\varepsilon = \pm 1$. In the following, we suppose without loss of generality that $\varepsilon = 1$ (otherwise, we can redefine $s_l(n)$ as its opposite).

Besides, the $k$th observation reads

$$x_k(n) = \sum_{q=1}^N a_{kq} s_q(n). \quad (83)$$

Let us now compute the differential correlation between $y(n)$ and $x_k(n)$, defined by

$$\mathrm{Dcorr}_{yx_k}(n_1, n_2) = E\{y(n_2)x_k(n_2)\} - E\{y(n_1)x_k(n_1)\}. \quad (84)$$

Then

$$\mathrm{Dcorr}_{yx_k}(n_1, n_2) = a_{kl}\mathrm{Dcorr}_{s_l s_l}(n_1, n_2) \\ + \sum_{q \neq l} a_{kq}\mathrm{Dcorr}_{s_l s_q}(n_1, n_2) \\ + \sum_{q_1 \notin I, q_2} \alpha_{q_1} a_{kq_2}\mathrm{Dcorr}_{s_{q_1} s_{q_2}}(n_1, n_2) \\ = a_{kl}\mathrm{Dpow}_{s_l}(n_1, n_2) = a_{kl} \quad (85)$$

as $\mathrm{Dcorr}_{s_{q_1}s_{q_2}}(n_1,n_2) = 0, \forall q_1 \neq q_2$ because of the independence of the zero-mean sources, $\mathrm{Dcorr}_{s_q s_q}(n_1, n_2) = 0, \forall q \notin I$ because of the stationarity of the noise sources and taking into account (82). Therefore, $\mathrm{Dcorr}_{yx_k}(n_1, n_2)$ yields the scale factor associated to the contribution of the $l$th source in the $k$th observation.

## APPENDIX C
## DIFFERENTIAL WIENER FILTERING

Let us consider the difference $x'_k(n)$ between observation $x_k(n)$ and a filtered version of the previously extracted signal $e_l(n)$ with $l \in I$, i.e.,

$$x'_k(n) = x_k(n) - \sum_{r=-R'}^{R'} c(r)e_l(n-r) \quad (86)$$

where $c(r)$ are the coefficients of a noncausal FIR filter. Due to the hypotheses of Section III-A, $x_k(n)$ is a causal FIR mixture of all innovation processes, i.e., (47) yields

$$x_k(n) = \sum_{q=1}^N \sum_{r=0}^L h_{kq}(r)u_q(n-r). \quad (87)$$

Moreover

$$e_l(n) = \beta_l u_l(n - r_l) + \sum_{q \notin I} \sum_p \beta_{qp} u_q(n-p) \quad (88)$$

where the bounds on $p$ need not be detailed here as the corresponding terms eventually vanish below in (92). Therefore, if $R'$ is high enough so that $r_l - R' \leq 0$ and $r_l + R' \geq L$, (86)–(88) yield

$$x'_k(n) = \sum_{r=r_l-R'}^{r_l+R'} [h_{kl}(r) - \beta_l c(r - r_l)]u_l(n-r) \\ + \sum_{q \in I, q \neq l} \sum_{r=0}^L h_{kq}(r)u_q(n-r) \\ + \sum_{q \notin I} \sum_p \gamma_{kq}(p)u_q(n-p) \quad (89)$$

where the weights $\gamma_{kq}(p)$ depend on $c(r), h_{kq}(r)$, and $\beta_{qp}$. These terms and the bounds on $p$ need not be detailed here for the same reason as previously. The differential power of $x'_k(n)$ is defined as

$$\mathrm{Dpow}_{x'_k}(n_1, n_2) = E\{x'_k(n_2)^2\} - E\{x'_k(n_1)^2\}. \quad (90)$$

Taking into account that all innovation processes are zero-mean, mutually and temporally independent, and that

$$\mathrm{Dpow}_{u_q}(n_1 - p, n_2 - p) = 0, \qquad \forall p \quad \forall q \notin I \quad (91)$$

equations (89) and (90) yield

$$\mathrm{Dpow}_{x'_k}(n_1, n_2) \\ = \sum_{r=r_l-R'}^{r_l+R'} [h_{kl}(r) - \beta_l c(r-r_l)]^2 \mathrm{Dpow}_{u_l}(n_1-r, n_2-r) \\ + \sum_{q \in I, q \neq l} \sum_{r=0}^L h_{kq}^2(r)\mathrm{Dpow}_{u_q}(n_1-r, n_2-r). \quad (92)$$

Since we assumed all differential powers $\mathrm{Dpow}_{u_l}(n_1-r, n_2-r)$ with $l \in I$ to be strictly positive, (92) shows that the minimization of $\mathrm{Dpow}_{x'_k}(n_1, n_2)$ with respect to $c(r)$ corresponds exactly to

$$h_{kl}(r) - \beta_l c(r-r_l) = 0, \quad \forall r \in [r_l - R', \ldots, r_l + R']. \quad (93)$$

By comparing this result to (89), we conclude that a criterion for obtaining a modified observation $x'_k(n)$ where the contributions of the $l$th source have been removed consists in minimizing $\mathrm{Dpow}_{x'_k}(n_1, n_2)$. We use Newton's method to optimize this criterion, which is denoted $J$ as follows and which may be expressed as

$$J = E\left\{\left(x_k(n_2) - \sum_{r=-R'}^{R'} c(r)e_l(n_2-r)\right)^2\right\}$$
$$- E\left\{\left(x_k(n_1) - \sum_{r=-R'}^{R'} c(r)e_l(n_1-r)\right)^2\right\}. \quad (94)$$

The update equation for the vector $\mathbf{c} = [c(-R'), \ldots, c(R')]^T$ reads

$$\mathbf{c} \leftarrow \mathbf{c} - \left(\frac{\partial^2 J}{\partial \mathbf{c}^2}\right)^{-1} \frac{\partial J}{\partial \mathbf{c}} \quad (95)$$

where $(\partial J/\partial \mathbf{c})$ and $(\partial^2 J/\partial \mathbf{c}^2)$ are, respectively, the gradient and the Hessian of the criterion $J$. Let us compute the $i$th component of the gradient of $J$ with respect to $\mathbf{c}$

$$\frac{\partial J}{\partial c_i} = 2E\left\{-e_l(n_2-i)\cdot\left(x_k(n_2) - \sum_{r=-R'}^{R'} c(r)e_l(n_2-r)\right)\right\}$$
$$- 2E\left\{-e_l(n_1-i)\cdot\left(x_k(n_1) - \sum_{r=-R'}^{R'} c(r)e_l(n_1-r)\right)\right\}. \quad (96)$$

In particular

$$\left.\frac{\partial J}{\partial c_i}\right|_{\mathbf{c}=0} = 2E\{-e_l(n_2-i)x_k(n_2)\} - 2E\{-e_l(n_1-i)x_k(n_1)\}. \quad (97)$$

Therefore, we have

$$\left.\frac{\partial J}{\partial \mathbf{c}}\right|_{\mathbf{c}=0} = -2\left(\mathbf{r}_{e_l x_k}(n_2) - \mathbf{r}_{e_l x_k}(n_1)\right) \quad (98)$$

where $\mathbf{r}_{e_l x_k}(n_p)$ is the cross-correlation vector between the signals $e_l(n_p - i)$, with $i \in [-R', R']$ and $x_k(n_p)$, with $p = 1$ or 2. Let us now compute the $(i,j)$th component of the Hessian of $J$

$$\frac{\partial^2 J}{\partial c_i \partial c_j} = 2E\{e_l(n_2-i)e_l(n_2-j)\}$$
$$- 2E\{e_l(n_1-i)e_l(n_1-j)\}. \quad (99)$$

Then, we have

$$\frac{\partial^2 J}{\partial \mathbf{c}^2} = 2\left(\mathbf{R}_{e_l}(n_2) - \mathbf{R}_{e_l}(n_1)\right) \quad (100)$$

where $\mathbf{R}_{e_l}(n_p)$ is the correlation matrix associated to the values of $e_l(n)$ around $n = n_p$. By initializing $\mathbf{c}$ with $\mathbf{c} = 0$, the value of $\mathbf{c}$ after one iteration of Newton's algorithm reads

$$\mathbf{c}' = -\left(\frac{\partial^2 J}{\partial \mathbf{c}^2}\right)^{-1} \left.\frac{\partial J}{\partial \mathbf{c}}\right|_{\mathbf{c}=0}$$
$$= -\frac{1}{2}(\mathbf{R}_{e_l}(n_2) - \mathbf{R}_{e_l}(n_1))^{-1}(-2)(\mathbf{r}_{e_l x_k}(n_2) - \mathbf{r}_{e_l x_k}(n_1))$$
$$= (\mathbf{R}_{e_l}(n_2) - \mathbf{R}_{e_l}(n_1))^{-1}(\mathbf{r}_{e_l x_k}(n_2) - \mathbf{r}_{e_l x_k}(n_1)). \quad (101)$$

Knowing that Newton's algorithm reaches a stationary point in one iteration for quadratic criteria and that our criterion is quadratic positive, we prove that the vector $\mathbf{c}'$ corresponds to the global minimum of $J$. Note that this expression is the differential version of the standard Wiener filter defined in (58).

## APPENDIX D
### DEFINITIONS OF $\text{SIR}_{\text{out}}$ AND $\text{SIR}_{\text{in}}$

Here, we first define the criteria used in Section IV to measure the performance of our deflation-based convolutive BSS method for underdetermined mixtures (the definitions for its instantaneous version follow). In each test, we first apply the mixtures of all sources to our BSS system and estimate its parameters, i.e., $\tilde{\mathbf{B}}, \mathbf{w}_k$ and $C_{ik}(z)$. Then, we freeze these parameters. Since we aim at evaluating the quality of the partial separation only achieved between the nonstationary sources, we then only transfer these sources through the mixing stage and the previously defined BSS system. Thus, we obtain a set of partial extracted signals $y'_k(n)$, with $k = 1, \ldots, P$. For each of them, we compute its colored versions $c_{ik}(n) * y'_k(n)$, respectively, associated to each observation $x_i(n)$. These colored signals are denoted $\hat{x}_{ik}(n)$ as follows. For each source $s_j(n)$ and each signal $\hat{x}_{ik}(n)$, we then define the associated $\text{SIR}_{\text{out}}(i,j,k)$ as the ratio of the "signal" and "interference" power as follows:

- the "signal" is the ideal value of $\hat{x}_{ik}(n)$ when it extracts $s_j(n)$, which is equal to the contribution of $s_j(n)$ in $x_i(n)$, that we denote $x_{ij}(n)$;
- the "interference" is the deviation of $\hat{x}_{ik}(n)$ from its ideal value, i.e., $\hat{x}_{ik}(n) - x_{ij}(n)$.

This yields $\forall i \in \{1, \ldots, P\}, \forall j \in I, \forall k \in \{1, \ldots, P\}$

$$\text{SIR}_{\text{out}}(i,j,k) = 10\log_{10}\left(\frac{E\{x_{ij}^2\}}{E\{(\hat{x}_{ik} - x_{ij})^2\}}\right). \quad (102)$$

For each source $j$ and output $k$, we only consider the single $\text{SIR}_{\text{out}}(i,j,k)$ corresponding to observation $i$ providing the signal $\hat{x}_{ik}$ which has the highest power. Then, for each source index $j$, we only consider the maximum with respect to output index $k$ of the values $\text{SIR}_{\text{out}}(i,j,k)$. Then, we derive the mean of these values on both time domains. This yields a single SIR for each of the sources of interest, defining the output performance of our system, which is denoted $\text{SIR}_{\text{out}}(j)$ hereafter. The mean over all sources of these $\text{SIR}_{\text{out}}(j)$ may then be derived. We perform this mean when computing the overall $\text{SIR}_{\text{out}}$ of symmetric methods.

The input SIR available from the observations is defined similarly as follows. First, we define the input SIR associated to each source as

$$\forall j \in I, \quad \text{SIR}_{\text{in}}(j) = \max_i 10\log_{10}\left(\frac{E\{x_{ij}^2\}}{\sum_{k \in I}^{k \neq j} E\{x_{ik}^2\}}\right). \quad (103)$$

Then, we derive the global $\text{SIR}_{\text{in}}$ as the mean of the previous contributions over all sources and both time domains.

## ACKNOWLEDGMENT

The authors would like to thank J. Chappuis for her participation in the early stages of this investigation.

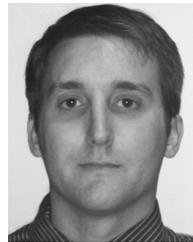

**Johan Thomas** was born in Quimper, France, in 1979. In 2004, he graduated from the Ecole Nationale de l'Aviation Civile, Toulouse, France and received the M.Sc. degree in electronics from the Institut National Polytechnique de Toulouse, Toulouse, France. Currently, he is working towards the Ph.D. degree at the Laboratoire d'Astrophysique de Toulouse-Tarbes which is part of the French National Center for Scientific Research.

He has been teaching at the Université Paul Sabatier Toulouse 3, Toulouse, France. His current research activities concern statistical signal processing, higher order statistics, neural networks, and especially, BSS methods with applications to astrophysics and acoustics.

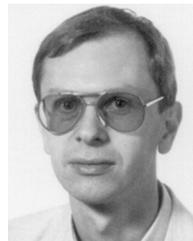

**Yannick Deville** (M'90) was born in Lyon, France, in 1964. He graduated from the Ecole Nationale Supérieure des Télécommunications de Bretagne, Brest, France, in 1986. He received the D.E.A and Ph.D. degrees, both in microelectronics, from the University of Grenoble, Grenoble, France, in 1986 and 1989, respectively.

From 1986 to 1997, he was a Research Scientist at Philips Research Labs, Limeil, France. His investigations during this period concerned various fields, including GaAs integrated microwave RC active filters, VLSI cache memory architectures and replacement algorithms, neural network algorithms and applications, and nonlinear systems. Since 1997, he has been a Professor at the University of Toulouse, Toulouse, France. From 1997 to 2004, he was with the Acoustics Lab of that University. Since 2004, he has been with the Astrophysics Lab in Toulouse, which is part of the University but also of the French National Center for Scientific Research (CNRS) and of the Midi-Pyrénées Observatory. His current major research interests include signal processing, higher order statistics, time-frequency analysis, neural networks, and especially, BSS and ICA methods and their applications to astrophysics, acoustics, and communication/electromagnetic signals.

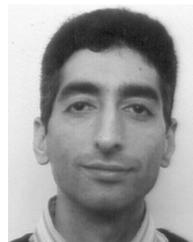

**Shahram Hosseini** was born in Shiraz, Iran, in 1968. He received the B.Sc. and M.Sc. degrees in electrical engineering from Sharif University of Technology, Tehran, Iran, in 1991 and 1993, respectively, and the Ph.D. degree in signal processing from the Institut National Polytechnique, Grenoble, France, in 2000.

He is currently an Associate Professor at the Université Paul Sabatier Toulouse 3, Toulouse, France. His research interests include BSS, artificial neural networks, and adaptive signal processing.